\documentclass[aps,pre,amsmath,superscriptaddress,amssymb,twocolumn]{revtex4}     
\usepackage{graphicx}

\begin{document}
\title{Free energy surface of ST2 water near the liquid-liquid phase transition}

\author{Peter H. Poole}
%\email{ppoole@stfx.ca}
\affiliation{Department of Physics, St. Francis Xavier University,
Antigonish, NS, B2G 2W5, Canada}

\author{Richard K. Bowles}
\affiliation{Department of Chemistry, University of Saskatchewan, Saskatoon, SK, 57N 5C9, Canada}

\author{Ivan Saika-Voivod}
\affiliation{Department of Physics and Physical Oceanography,
Memorial University of Newfoundland, St. John's, NL, A1B 3X7, Canada}

\author{Francesco Sciortino}
\affiliation{Dipartimento di Fisica,
Universit\`a di Roma {\em La Sapienza},
Piazzale A. Moro 5, 00185 Roma, Italy}

\begin{abstract}
We carry out umbrella sampling Monte Carlo simulations to evaluate the free energy surface of the ST2 model of water as a function two order parameters, the density and a bond-orientational order parameter.  We approximate the long-range electrostatic interactions of the ST2 model using the reaction-field method.  We focus on state points in the vicinity of the liquid-liquid critical point proposed for this model in earlier work.  At temperatures below the predicted critical temperature we find two basins in the free energy surface, both of which have liquid-like bond orientational order, but differing in density.  The pressure and temperature dependence of the shape of the free energy surface is consistent with the assignment of these two basins to the distinct low density and high density liquid phases previously predicted to occur in ST2 water.
\end{abstract}

\date{\today}
\maketitle

\section{Introduction}

\begin{figure}\bigskip\bigskip
\centerline{\includegraphics[scale=0.35]{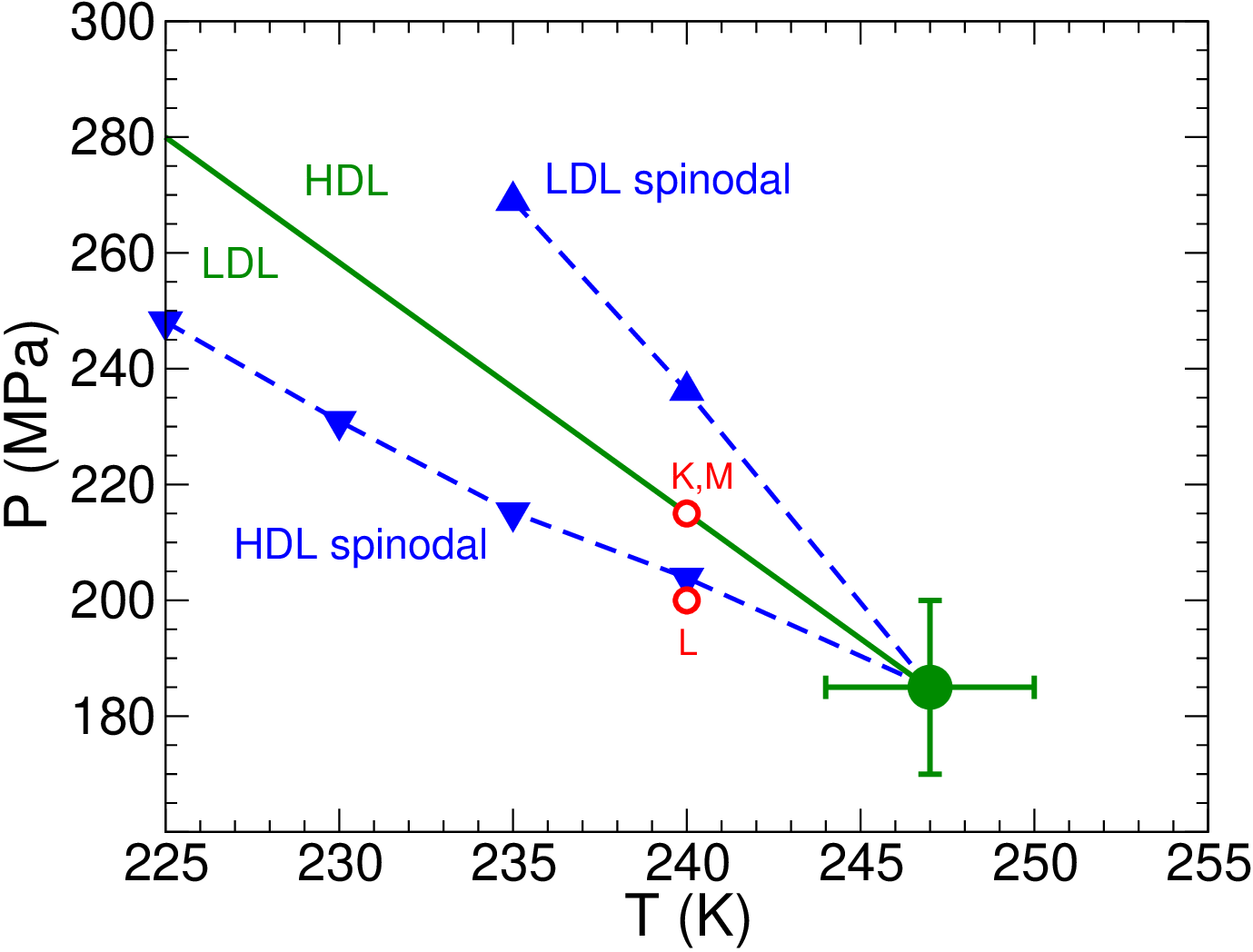}}
\caption{Phase behavior of the ST2-RF model predicted from previous work using $N=1728$ molecular dynamics simulations.  Shown are the estimated locations of the critical point (green circle) and the LDL-HDL coexistence line (green line)~\cite{cuth}.  
Note that the error bars associated with the critical point also apply to the coexistence line.
Estimates for the HDL spinodal (down-triangles) and LDL spinodal (up-triangles) are also shown~\cite {denmin}.  Red circles locate the state points at which we carry out series K, L, and M of the present work.}
\label{map}
\end{figure}

In 1992, the results of a computer simulation study of the ST2 model~\cite{st2} of water were used to propose that a liquid-liquid phase transition (LLPT) occurs in supercooled water~\cite{PSES}.  Below the critical temperature $T_c$ for the proposed LLPT, two distinct phases of water, the low density liquid (LDL) and high density liquid (HDL) phases are separated by a first-order phase transition.  The predicted phase diagram for the ST2 model in the plane of temperature $T$ and pressure $P$ in the vicinity of the critical point is shown in Fig.~\ref{map}.  

An appealing feature of the LLPT proposal is that it simultaneously accounts for (a) the unusual thermodynamic behavior of liquid water in the supercooled region, and (b) the occurrence of two distinct forms of amorphous solid water in the glassy regime~\cite{stanley,pdreview}.  Evidence for a LLPT has been reported in a number of simulation studies of water and water-like systems; see e.g.~\cite{harr,denmin,paschek,sastry,sri,vega,jagla}.  Experimentally, a LLPT has yet to be decisively confirmed in supercooled water, and efforts to resolve this question in the laboratory continue~\cite{wink,huang,clark,zhang}.  The predicted location of the critical point in the supercooled regime is challenging to study in experiments because of rapid ice crystallization.  In simulations, this problem is avoided when the liquid can be studied on a time scale that is long relative to the liquid-state relaxation time, but short compared to crystal nucleation times.

Recently, Limmer and Chandler~\cite{lim} have challenged the LLPT hypothesis.  Using umbrella sampling Monte Carlo (MC) simulations of two water models (mW~\cite{mw} and ST2 water), Ref.~\cite{lim} presents results for the free energy surface $F(\rho,Q_6)$ of the liquid as a function of two order parameters, the density $\rho$, and a bond-orientational order parameter $Q_6$.  $Q_6$ is a bulk order parameter used to distinguish crystalline configurations from liquid or amorphous solid states of a system.   Values of $Q_6$ approaching zero correspond to disordered states, while larger values of $Q_6$ indicate greater degrees of crystalline order.  The detailed definition of $Q_6$ is given in Eqs.~1-3 of Ref.~\cite{lim}, and is based on an analysis of the orientation of local molecular environments (i.e. a molecule and its nearest neighbors) in terms of spherical harmonics, as originally proposed by Steinhardt, et al.~\cite{q6}.  In the present work, we use the same definition of $Q_6$ as given in Ref.~\cite{lim}.

It has long been appreciated that the density of the proposed LDL and HDL phases must be different.  The innovation of Ref.~\cite{lim} is that by examining the dependence of $F(\rho,Q_6)$ on $Q_6$, Limmer and Chandler address the relationship of the metastable liquid phase to the ordered crystalline ice phases.  If a LLPT transition occurs in a simulation model, then under appropriate conditions of $T$ and $P$, two distinct free energy basins should be observed in $F(\rho,Q_6)$ in the low-$Q_6$ (i.e. liquid-like) regime.  For both the mW and ST2 water models, Ref.~\cite{lim} reports that only one liquid-like free energy basin is found in $F(\rho,Q_6)$, including, in the case of ST2 water, at conditions below the proposed critical temperature of the LLPT.  Limmer and Chandler conclude that phenomena previously interpreted as evidence for a LLPT are in fact due to the liquid-to-crystal phase transition.

Since the publication of Ref.~\cite{lim}, Liu et al. have reported on their own evaluation of the free energy surface $F(\rho,Q_6)$ found from umbrella sampling MC simulations of ST2 water~\cite{liu}.  Although they employ methods similar to those used in Ref.~\cite{lim}, Liu et al. report a very different result:  the observation of two distinct liquid free energy basins in $F(\rho,Q_6)$, with properties consistent with the LLPT hypothesis.  The results of Ref.~\cite{liu} are also consistent with an earlier study by the same group reporting the free energy of ST2 water as a function of $\rho$ only~\cite{liu1}.  

\begin{figure}\bigskip\bigskip
\centerline{\includegraphics[scale=0.35]{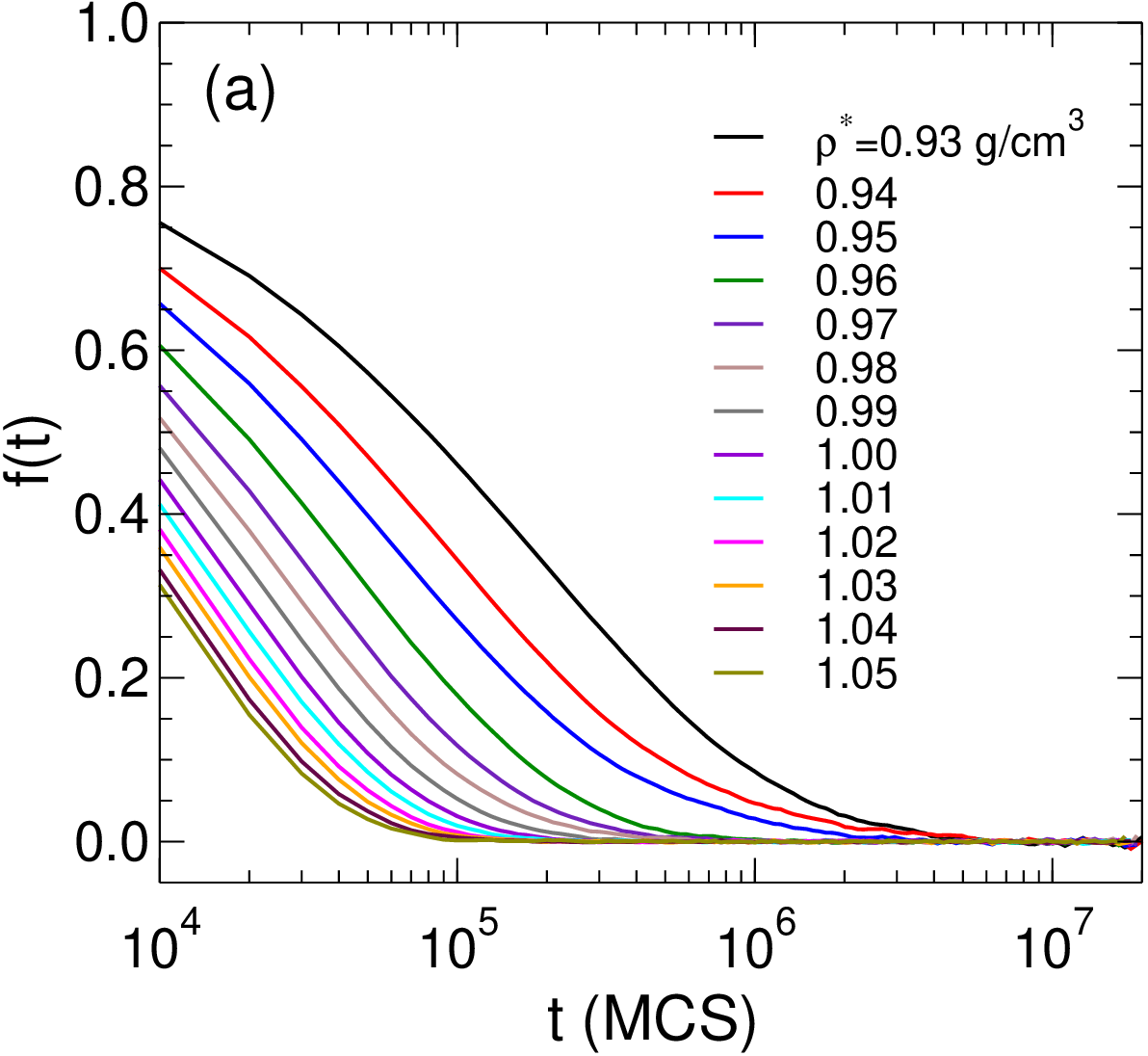}}
\bigskip\bigskip
\centerline{\includegraphics[scale=0.35]{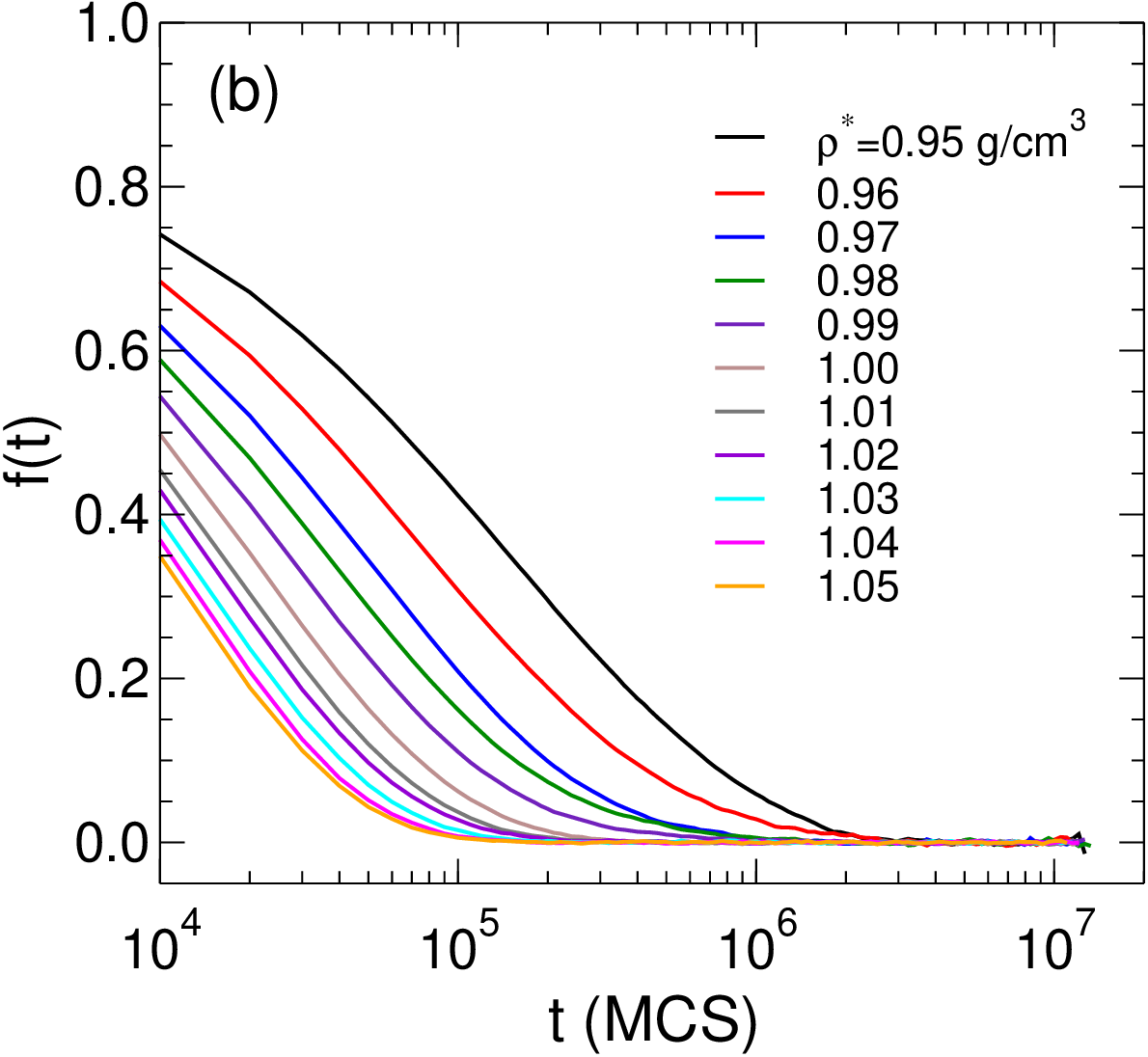}}
\caption{Time dependence of the collective intermediate scattering function $f(t)$ for runs with various values of $\rho^*$ in series (a) K and (b) L.  Each curve is an average over 10 runs.}
\label{fqt-all}
\end{figure}

The precise reasons for the difference between the results of Refs.~\cite{lim} and \cite{liu} for $F(\rho,Q_6)$ remain unclear.  Among the differences in the approaches used in these two works, we note two.  First, Limmer and Chandler present results for $F(\rho,Q_6)$ at various pressures as determined at one temperature, $T=235$~K, which is below but within error of the estimated critical temperature $T_c=237\pm 4$~K for the ST2 model when studied with Ewald summations~\cite{liu1}.  Working this close to $T_c$ may make it difficult to discern distinct liquid basins in the free energy surface within the statistical error.  Liu et al. report results for a range of temperatures below $T_c$, from $T=224$ to $235$~K, and show that the distinction between the two liquid basins that they observe in $F(\rho,Q_6)$ becomes greater as $T$ decreases below $T_c$.

Second, in both Refs.~\cite{lim} and \cite{liu}, the method of Ewald summation is used to approximate the long-range contributions to the electrostatic potential energy of the ST2 system.  However, Liu et al. report that their Ewald summation method employs vacuum boundary conditions, whereas Limmer and Chandler use conducting boundary conditions.  Liu et al. note some significant sensitivity in the behavior of their system as a function of these boundary condition choices.  If and how these boundary conditions might affect the qualitative shape of the $F(\rho,Q_6)$ surface is incompletely understood.  

In light of the conflicting results of Refs.~\cite{lim} and \cite{liu}, we present here a new evaluation of the free energy surface $F(\rho,Q_6)$ of ST2 water.  In order to expand our understanding of the role of long-range interactions,  we use a different approach to account for the electrostatic energy, namely the reaction field method~\cite{rf}.  Indeed, many of the previous studies of ST2 water that relate to the LLPT hypothesis were conducted using the reaction field method~\cite{harr,denmin,sus,kessel}, including the work in which the occurrence of a LLPT was first proposed~\cite{PSES}.  Furthermore, a recent umbrella sampling MC study of the ST2 model, using the reaction field method, showed that the shape of the free energy as a function of $\rho$ was consistent with the LLPT hypothesis~\cite{sus}.  An explicit examination of the $F(\rho,Q_6)$ surface for the ST2 model with a reaction field treatment of the electrostatics is therefore warranted.  In addition, we also study a range of temperatures and pressures in the vicinity of the proposed critical point, to examine their influence on the $F(\rho,Q_6)$ surface.

\begin{figure*}\bigskip
\centerline{\includegraphics[scale=0.6]{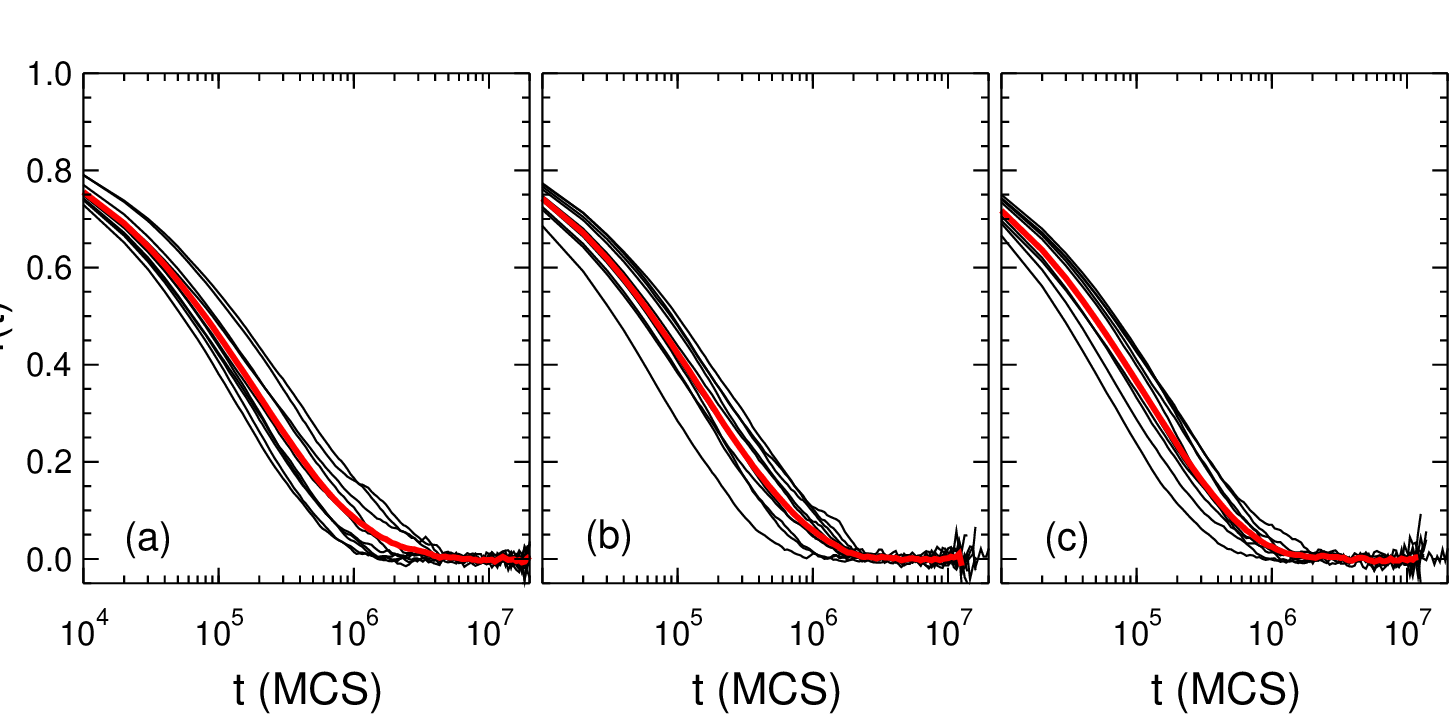}}
\caption{Collective intermediate scattering function $f(t)$ for the lowest value of $\rho^*$ from each series: (a) K, $\rho^*=0.93$~g/cm$^3$; (b) L, $\rho^*=0.95$~g/cm$^3$; and (c) M, $\rho^*=0.95$~g/cm$^3$.  These are the most slowly relaxing runs used in our analysis.  The black lines give $f(t)$ for each of the 10 independent runs conducted at the same values of $(T,P,\rho^\ast,Q_6^\ast)$, and the thick red line is their average.}
\label{fqt}
\end{figure*}

\section{ST2 model}

We study the ST2 model of water proposed by Stillinger and Rahman ~\cite{st2}.  The ST2 pair potential is a sum of a Lennard-Jones (LJ) interaction (centered on the O atom), and electrostatic interactions involving four tetrahedrally positioned charges.  Our model parameters for the geometry and pair interactions of the ST2 water molecule are the same as those given in Ref.~\cite{st2}.  The potential energy $U$ of our system is given by,
\begin{equation}
U= U_{\rm E} + U_{\rm LJ} +\Delta U_{\rm LJ},
\label{U}
\end{equation}
where $U_{\rm E}$ and $U_{\rm LJ}$ are the respective electrostatic and LJ contributions.  In our simulations, the LJ interaction is sharply cut off when the O-O distance $r$ exceeds $R_c=0.78$~nm, and the contribution from longer ranged LJ interactions is approximated by, 
\begin{equation}
\Delta U_{\rm LJ}= - \frac{8\pi \epsilon \sigma^6\rho_n N}{3R_c^3},
\label{ULJ}
\end{equation}
as described in the Appendix of Ref.~\cite{st2}.  In Eq.~\ref{ULJ}, $N$ is the number of molecules, $\rho_n$ is the number density of molecules, and $\epsilon$ and $\sigma$ are the respective energy and size parameters of the LJ potential.

To evaluate $U_{\rm E}$, the electrostatic contributions to the potential energy, we adopt the treatment used in the study of ST2 water by Steinhauser; see Eqs.~5 and 6 of Ref.~\cite{rf}.  In this approach, the electrostatic interactions of the ST2 model are evaluated directly up to $r=R_c$ using the original form given in Ref.~\cite{st2}, including the use of a ``switching function" to preclude a divergence of the energy due to charge overlaps.  The contribution of electrostatic interactions beyond $R_c$ is then approximated using the reaction field method, in which the liquid beyond $R_c$ is treated as a polarizable dielectric continuum.  As in Ref.~\cite{rf}, we assume that the dielectric constant of the continuum liquid is $\epsilon_{\rm R}=\infty$.  To avoid a sharp discontinuity in the electrostatic interactions at $R_c$, a tapering function (described in Ref.~\cite{rf}) is used to smoothly reduce the electrostatic interaction between two molecules (both direct and reaction field contributions) to zero over the interval $0.95R_c<r<R_c$.  

The evaluation of the pair interactions as described above is the same procedure that was used in a number of previous studies~\cite{PSES,harr,denmin,cuth,becker,sus}.  For the remainder of this paper, we will refer to the reaction field version of ST2 described above as ST2-RF, to emphasize the difference between the present study and those works that have studied the ST2 model using an Ewald treatment of the electrostatics~\cite{liu1,lim,liu}.

\section{Simulation Methods}

Our aim is to evaluate the free energy surface $F(\rho,Q_6)$ for the ST2-RF model in the vicinity of the predicted LLPT for this model.  To define $F(\rho,Q_6)$, let $p(\rho,Q_6)$ be proportional to the equilibrium probability for a microstate of the system at fixed values of $N$, $T$, and $P$ to have order parameter values $\rho$ and $Q_6$.  The conditional Gibbs free energy $F(\rho,Q_6)$ is then defined by, 
\begin{equation}
F(\rho,Q_6)=-kT\ln p(\rho,Q_6) + F_0
\end{equation}
where $F_0$ is an (irrelevant) constant related to the normalization of $p$, and $k$ is Boltzmann's constant~\cite{lim}.

We also define the ``contraction" of $F$ with respect to $Q_6$ as,
\begin{equation}
{\bar F}(\rho)=-kT\ln \biggl (\int_0^{Q_6^{\rm max}}dQ_6\,\exp [-\beta F(\rho,Q_6)]\biggr),
\end{equation}
where $\beta=1/kT$~\cite{lim}.  ${\bar F}(\rho)$ represents the free energy as a function of $\rho$ that would be found from an ensemble of states in which $Q_6$ is free to vary between zero and $Q_6^{\rm max}$.  In this work, we are concerned with the liquid-like range of $Q_6$.  As shown below, we find that setting $Q_6^{\rm max}=0.09$ is sufficient to characterize ${\bar F}(\rho)$ for the liquid-like basins of the free energy surface.

\begin{figure}\bigskip
\centerline{\includegraphics[scale=0.38]{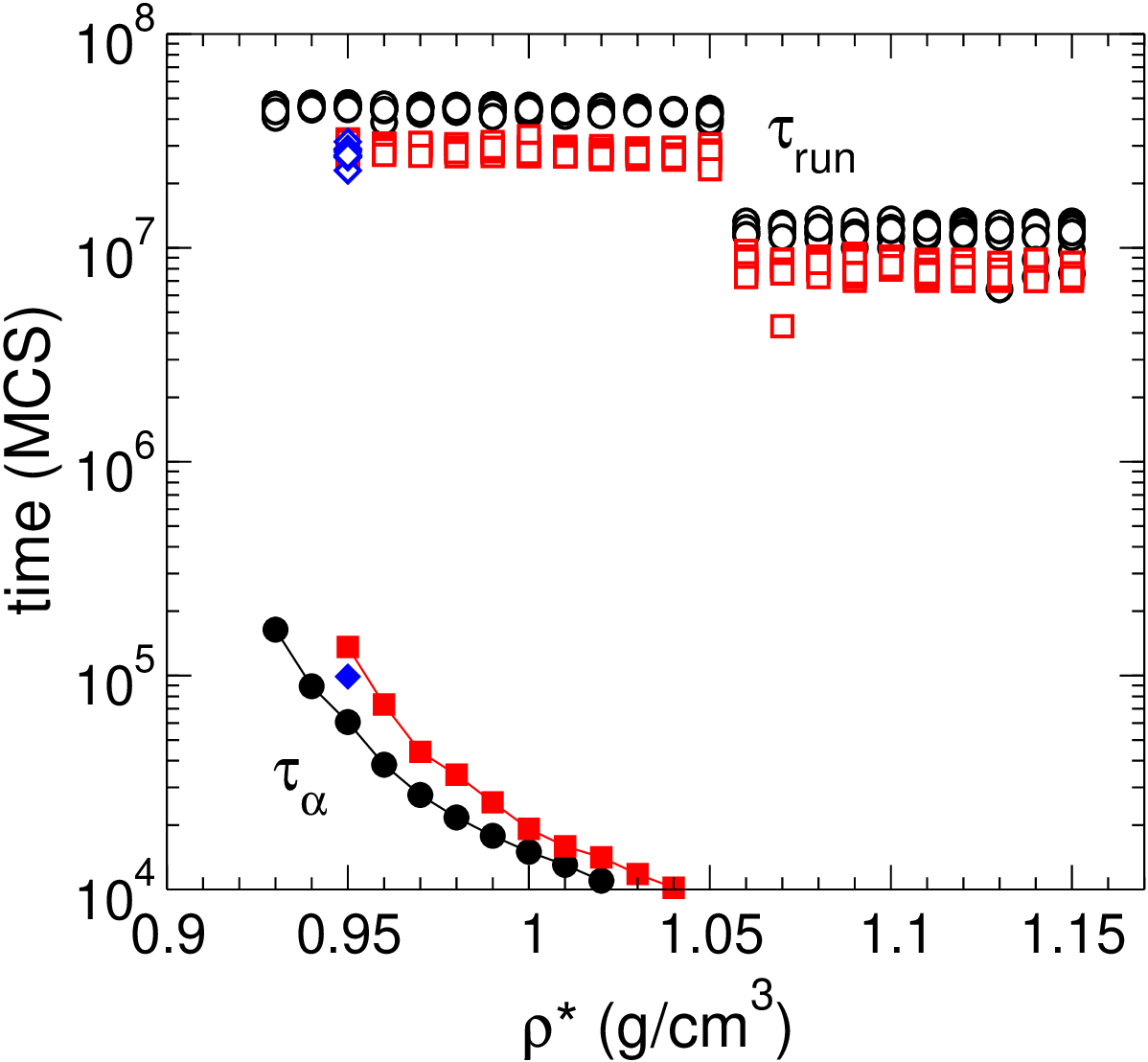}}
\caption{Comparion of $\tau_\alpha$ (filled symbols) and $\tau_{\rm run}$ (open symbols) as a function of $\rho^\ast$ for series K (circles), L (squares), and M (diamonds).  Values of $\tau_\alpha<10^4$~MCS are not shown.}
\label{times}
\end{figure}

\begin{figure}\bigskip
\centerline{\includegraphics[scale=0.38]{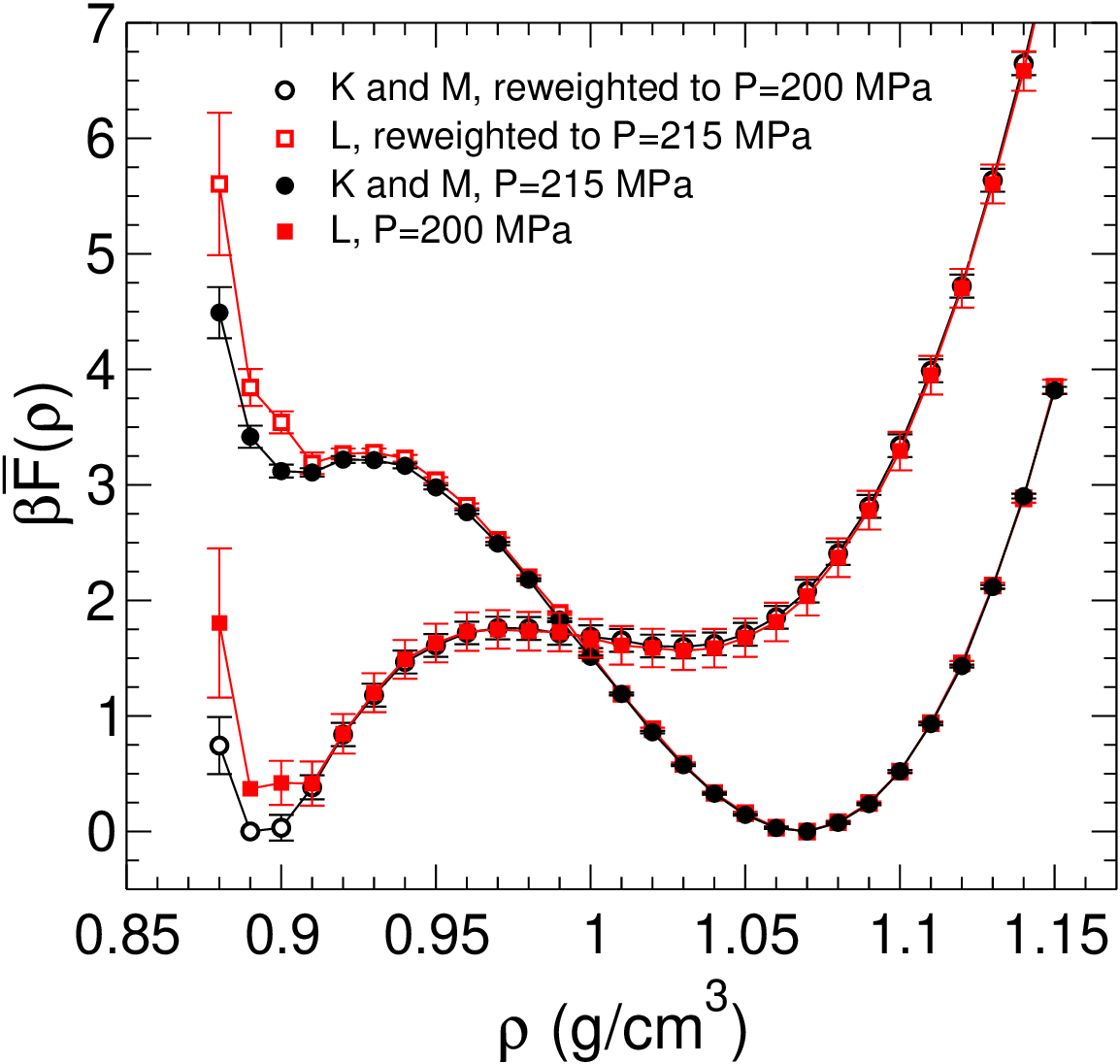}}
\caption{Contracted free energy ${\bar F}(\rho)$ at $T=240$~K, at two different pressures.  The filled circles are obtained by analyzing series K and M at $P=215$~MPa, the pressure at which these series are conducted;  the open circles are obtained by reweighting these results to $P=200$~MPa. The filled squares are obtained by analyzing series L at $P=200$~MPa, the pressure at which this series is conducted; the open squares are obtained by reweighting these results to $P=215$~MPa.}
\label{KLM}
\end{figure}

Following the approach of Ref.~\cite{lim}, we use umbrella sampling MC simulations 
to evaluate $F(\rho,Q_6)$.
We carry out MC simulations in the constant-$(N,P,T)$ ensemble, and to implement umbrella sampling, we add a biasing potential,
\begin{equation}
U_{\rm B}=k_1(\rho-\rho^\ast)^2+k_2(Q_6-Q_6^\ast)^2
\end{equation}
to the system potential energy $U$ in Eq.~\ref{U}.  The effect of $U_{\rm B}$ is to constrain a given simulation to sample configurations in the vicinity of chosen values of the order parameters $\rho=\rho^\ast$ and $Q_6=Q_6^\ast$.  In all our simulations, we fix $N=216$, $k_1=1000kT$~(cm$^3$/g)$^2$, and $k_2=2000kT$.

Trial configurations for each Monte Carlo step (MCS) are generated as follows:  First, we carry out a mini-trajectory of 10 unbiased (i.e. $U_{\rm B}=0$) constant-$(N,P,T)$ MC moves, in which each move consists (on average) of $N-1$ attempted rototranslational moves, and one attempted change of the system volume.  The maximum size of the attempted rototranslational and volume changes are chosen to give MC acceptance ratios in the range 25-40\%, depending on the thermodynamic conditions.  Next, the change in the biasing potential $U_{\rm B}$ is evaluated for the trial configuration resulting from the mini-trajectory, relative to the system configuration at the beginning of the mini-trajectory, to determine the acceptance or rejection of the trial configuration.  This completes one MCS, and the procedure is then repeated.

In order to identify the $T$-$P$ state points at which to conduct our runs, we use the location of the LLPT reported in previous work.  Fig.~\ref{map} shows the estimates for the critical point and coexistence line obtained from $N=1728$ molecular dynamics simulations of the ST2-RF model.  Of particular importance are the locations of the spinodal lines for the LDL and HDL phases.  These spinodal lines demarcate the stability limits for each phase.  Consequently, if liquid-liquid coexistence does indeed occur in the ST2-RF model, the $F(\rho,Q_6)$ surface will simultaneously exhibit two distinct liquid basins only for state points lying in the region between the HDL and LDL spinodals.  It is in this region of states that we focus our simulations.  To carry out our runs, we select pressures that lie between or near to the HDL and LDL spinodals, and a temperature ($T=240$~K) that is 7~K below the estimated critical temperature of $T_c=247\pm 3$ for the ST2-RF model~\cite{cuth}.

We carry out three distinct series of runs.  In the following, ``series K" denotes the set of runs conducted at  $T=240~{\rm K}, P=215~{\rm MPa}$, $Q_6^\ast=0.05$, and equally spaced values of $\rho^\ast$ from 0.93 to 1.15~g/cm$^3$, separated by 0.01~g/cm$^3$.  ``Series L" denotes runs conducted at $T=240~{\rm K}, P=200~{\rm MPa}$, $Q_6^\ast=0.05$, and equally spaced values of $\rho^\ast$ from 0.95 to 1.15~g/cm$^3$, separated by 0.01~g/cm$^3$.  ``Series M"  denotes runs conducted at $T=240~{\rm K}, P=215~{\rm MPa}$, $Q_6^\ast=0.09$, and $\rho^\ast=0.95$~g/cm$^3$.  The state points in the $T$-$P$ plane corresponding to series K, L, and M are identified in Fig.~\ref{map}.  For all distinct choices of $(T,P,\rho^\ast,Q_6^\ast)$ in the above series, we conduct 10 separate runs, each initiated from independent starting configurations.  The results presented here are thus based on an analysis of 450 independent runs.

\begin{figure}\bigskip
\centerline{\includegraphics[scale=0.38]{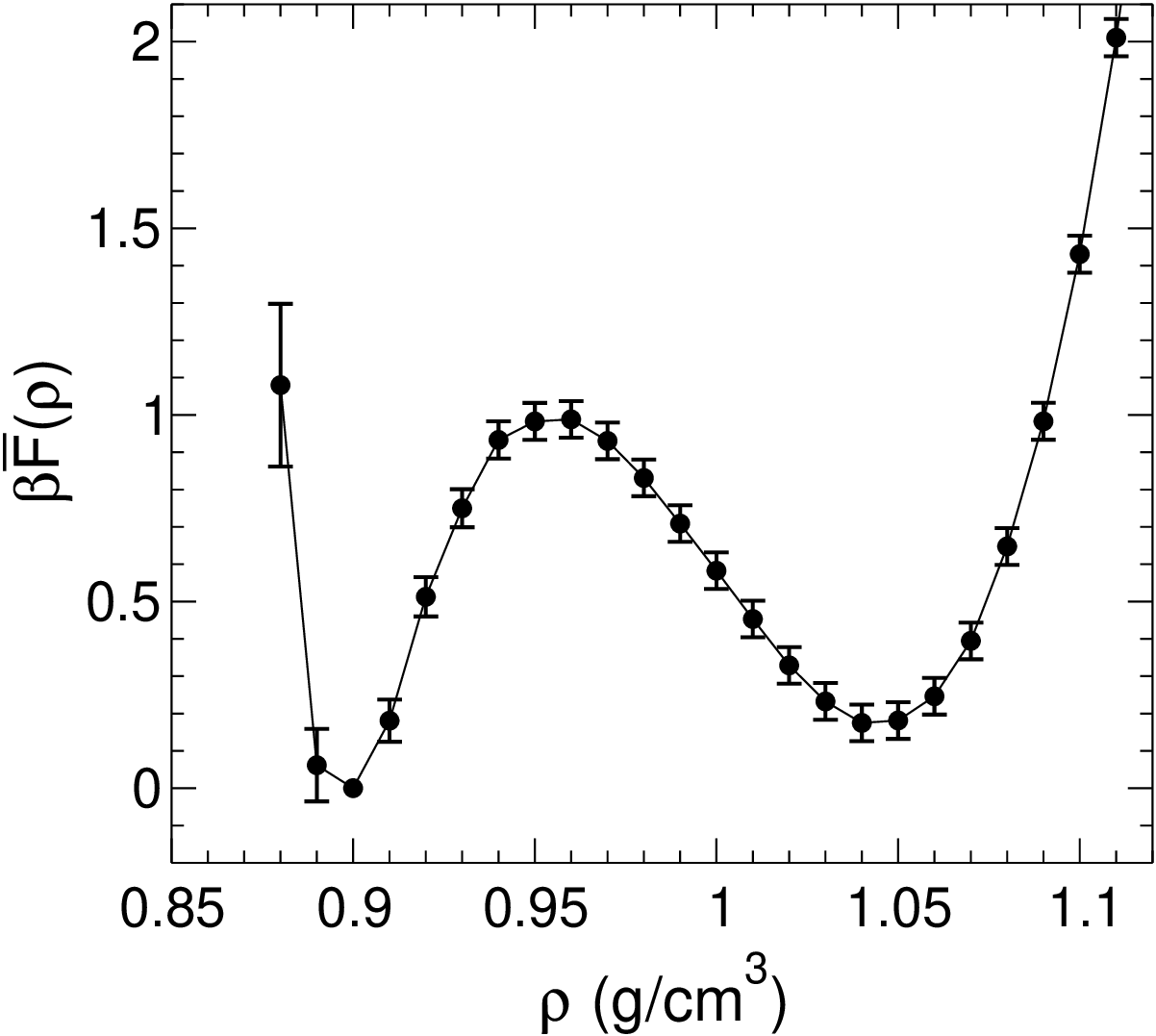}}
\caption{Contracted free energy ${\bar F}(\rho)$ at $T=240$~K and $P=204.5$~MPa, obtained by combining all results from series K, L, and M.}
\label{P204}
\end{figure}

All our runs are carried out for between $5\times 10^6$ to $5\times 10^7$ MCS.  Using the second half of each run, we compute $f(t)$, the collective intermediate scattering function as a function of time $t$.  We evaluate $f(t)$ at the lowest-wavenumber peak in the static structure factor for the O atoms, i.e. the so-called first sharp diffraction peak of molecular tetrahedral networks.  As shown in Figs.~\ref{fqt-all} and \ref{fqt}, in all cases $f(t)$ decays to zero on a time scale which is short compared to the lengths of our runs.  Hence the system behaviour is consistent with liquid-like relaxation under all conditions simulated in this study.  After averaging $f(t)$ over the 10 runs at each choice of $(T,P,\rho^\ast,Q_6^\ast)$, we estimate the alpha-relaxation time $\tau_\alpha$ as the time at which $f(t)=e^{-1}$.  As shown in Fig.~\ref{times}, in all cases we find $\tau_\alpha<2\times 10^5$~MCS.  To account for equilibration, we then discard the results for $t<\tau_e$ of each run, where $\tau_e=20 \tau_\alpha$ or $10^4$~MCS, whichever is larger.  The resulting length $\tau_{\rm run}$ of each production run that is used in our analysis is shown in Fig.~\ref{times}, compared to the corresponding value of $\tau_\alpha$.  In terms of $\tau_\alpha$, the lengths of our production runs range between 175$\tau_\alpha$ and 4400$\tau_\alpha$.  As shown in Fig.~\ref{fqt}, even our most slowly relaxing individual simulations are run for a time that is at least two orders of magnitude longer than the corresponding value of $\tau_\alpha$.  

\begin{figure*}\bigskip
\centerline{\includegraphics[scale=0.6]{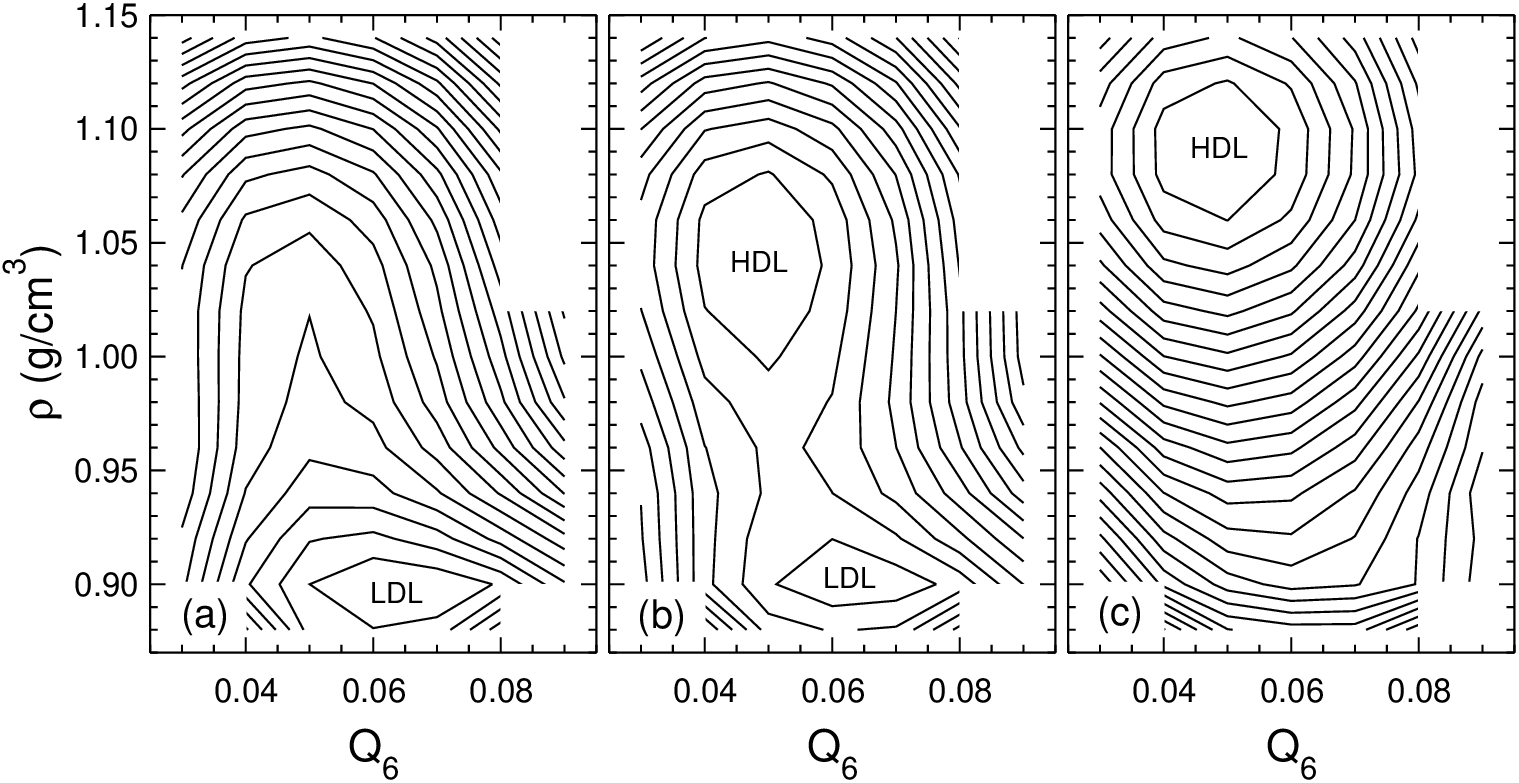}}
\caption{Contour plots of $F(\rho,Q_6)$ at $T=240$~K for (a) $P=195$~MPa, (b) $P=204.5$~MPa, and (c) $P=230$~MPa.  
To evaluate these surfaces, we have coarse-grained the plane of $\rho$ and $Q_6$ into rectangular cells of dimensions $\Delta \rho=0.02$~g/cm$^3$ and $\Delta Q_6=0.01$.  
Data from all series (K, L, and M) are combined and analyzed to obtain these plots.  For each panel, contours are separated by $0.5kT$, and the error in $F$ is $0.5kT$ or less.  The lowest lying values of $F$ in each plot are labelled LDL and/or HDL.}
\label{cont}
\end{figure*}

To estimate $F(\rho,Q_6)$, ${\bar F}(\rho)$, and the associated error, we use the multistage Bennet acceptance ratio (MBAR) method~\cite{mbar}.  The MBAR method takes as input the time series of the order parameters ($\rho$ and $Q_6$) and the system potential energy $U$, reweights the statistics obtained from each run to remove the effect of the biasing potential, and produces an optimal estimate of the desired free energy function at a specified value of $T$ and $P$.  The MBAR method also facilitates reweighting the configurations sampled during our runs with respect to $T$ and/or $P$, allowing the statistics from different state points to be combined to produce an estimate of $F(\rho,Q_6)$ or ${\bar F}(\rho)$ at $T$-$P$ state points that lie near to the conditions at which we carry out our simulations.

For the purpose of estimating the free energy and its error using MBAR, we wish to consider only those configurations from our runs that are statistically independent.  We assume that statistically independent configurations are separated by $\tau_\alpha$ or $10^4$~MCS, whichever is larger.  All other configurations are ignored in our analysis.  Note that in all our plots the indicated error is the error with respect to the minimum value of the estimated free energy, which in most cases is arbitrarily set to zero.  Also, all error bars reported here represent one standard deviation of error.

\section{Results}

First, we compare the results obtained for ${\bar F}(\rho)$ at the two state points directly simulated in our runs.  Series K and M are both conducted at $T=240$~K and $P=215$~MPa, while series L is conducted at $T=240$~K and $P=200$~MPa.  The results for ${\bar F}(\rho)$ obtained using only series K and M, and that obtained using only series L are compared in Fig.~\ref{KLM}.  The shapes of both curves suggest the existence of two distinct free energy minima separated by an interval of thermodynamic instability with respect to $\rho$, as indicated by concave-down curvature of ${\bar F}(\rho)$.   One minimum is centred near $0.9$~g/cm$^3$ and the other near $1.05$~g/cm$^3$.  

To check that the statistics we have gathered in series K and M are consistent with the results obtained from series L (and vice versa), we also show in Fig.~\ref{KLM} the result for ${\bar F}(\rho)$ found by reweighting our data from series K and M to $P=200$~MPa, and the result found by reweighting our data from series L to $P=215$~MPa.  The reweighted results are in good agreement with the unreweighted curves, confirming that both data sets have independently converged to equilibrium.  In the remainder of this paper, all results shown for ${\bar F}(\rho)$ and $F(\rho,Q_6)$ are therefore obtained by combining the statistics from all three simulations series, K, L, and M.

In Fig.~\ref{P204} we show ${\bar F}(\rho)$ at $T=240$~K and $P=204.5$~MPa, a pressure intermediate between those shown in Fig.~\ref{KLM}.  At this state point, ${\bar F}(\rho)$ clearly displays two distinct free energy minima, separated by a free energy barrier of approximately $1kT$, a typical value when $T$ is close to $T_c$.

\begin{figure}\bigskip
\centerline{\includegraphics[scale=0.34]{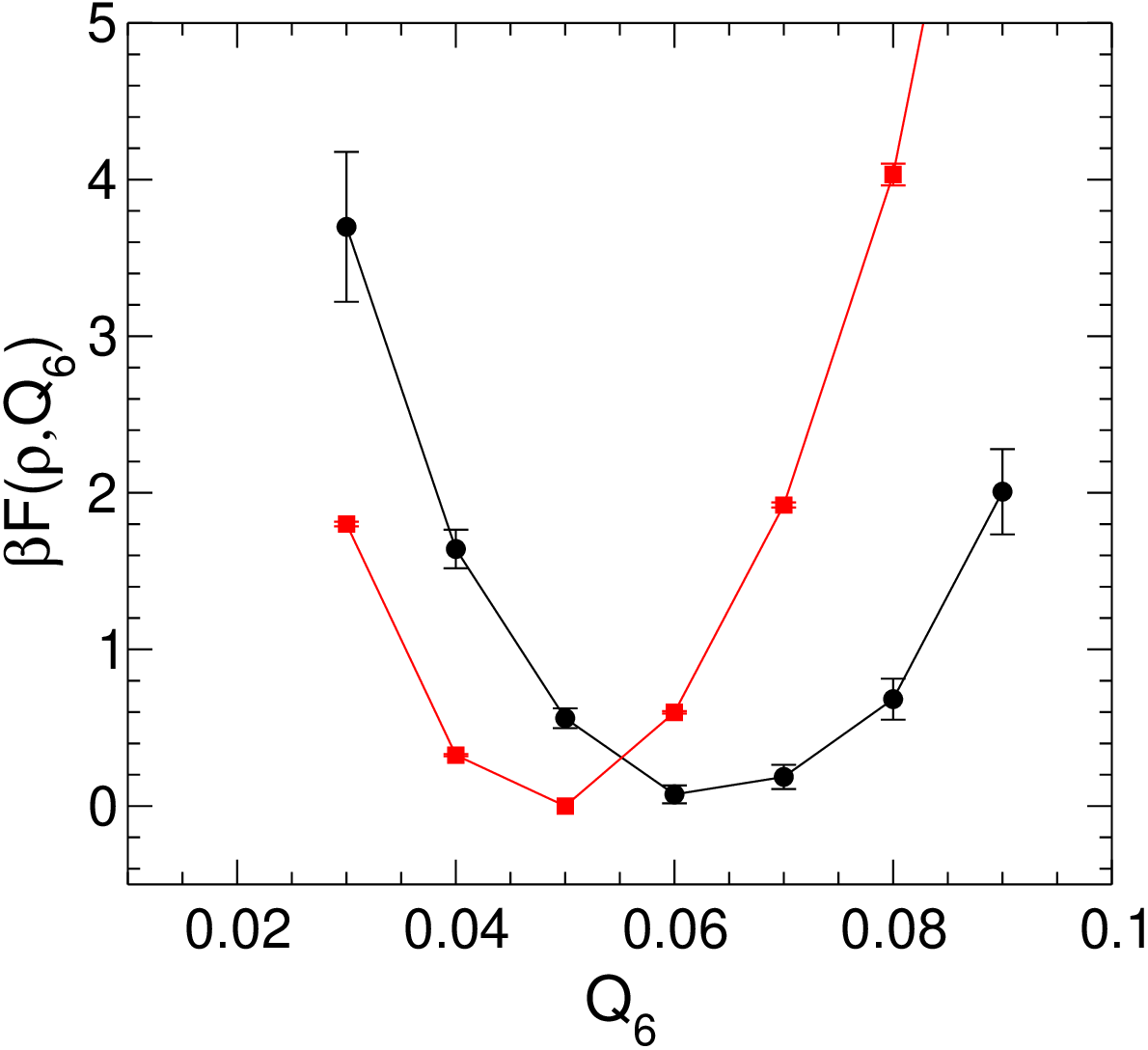}}
\caption{Slices through the free energy surface $F(\rho,Q_6)$ for $T=240$~K and  $P=204.5$~MPa [shown in Fig.~\ref{cont}(b)] as a function of $Q_6$, passing through the minima of the LDL basin at $\rho=0.90$~g/cm$^3$ (circles), and the HDL basin at $\rho=1.04$~g/cm$^3$ (squares).}
\label{q6}
\end{figure}

We next analyze the behavior of the free energy surface $F(\rho,Q_6)$.  Fig.~\ref{cont} shows contour plots of $F(\rho,Q_6)$ at $T=240$~K for three pressures from $P=195$ to $230$~MPa.  The $F(\rho,Q_6)$ surface at $P=204.5$~MPa simultaneously displays two free energy basins, each corresponding to a distinct metastable thermodynamic phase.  The minima of both basins are located at liquid-like values of $Q_6$ in the range $0.05$-$0.065$.  The shape of both basins shows that the phases they represent are locally stable with respect to fluctuations in both $\rho$ and $Q_6$.  The stability of both phases with respect to $Q_6$, highlighted in Fig.~\ref{q6}, shows that neither free energy basin is connected via a monatonic ``downhill" path to any of the free energy basins associated with the various phases of crystalline ice, which are expected to occur at much higher values of $Q_6\simeq 0.5$.  The properties of the phases associated with the two basins shown in Fig.~\ref{cont}(b) are therefore consistent with two distinct liquids, the LDL and HDL phases, predicted to occur in the ST2-RF model in earlier work~\cite{PSES,denmin,sus}.

If the two basins shown in Fig.~\ref{cont}(b) are consistent with a LLPT between LDL and HDL phases, then increasing the pressure at constant $T$ should cause the LDL basin to disappear, and decreasing the pressure should cause the HDL basin to disappear, as both phases reach the respective spinodal limits that bracket the coexistence curve (see Fig.~\ref{map}).  This is illustrated in Fig.~\ref{cont}(a) and (c).  At $P=195$~MPa, only the LDL basin remains, while at $P=230$~MPa only the HDL basin is observed.  

We note that at $T=240$~K, the pressure range found here that corresponds to the region between the HDL and LDL spinodals appears to be shifted downward by about 10~MPa relative to the thermodynamic features shown in Fig.~\ref{map}.  However, this difference is less than the error associated with the results in Fig.~\ref{map} for the location of the critical point and coexistence line.  Considering that the features in Fig.~\ref{map} are based on an extrapolation of equation-of-state data from $N=1728$ molecular dynamics simulations~\cite{denmin,cuth}, and considering the possibility of differences due to finite-size effects when comparing with our $N=216$ results, the agreement between the behavior observed here and that predicted in Fig.~\ref{map} is quite satisfactory.

\begin{figure}[t]\bigskip
\centerline{\includegraphics[scale=0.35]{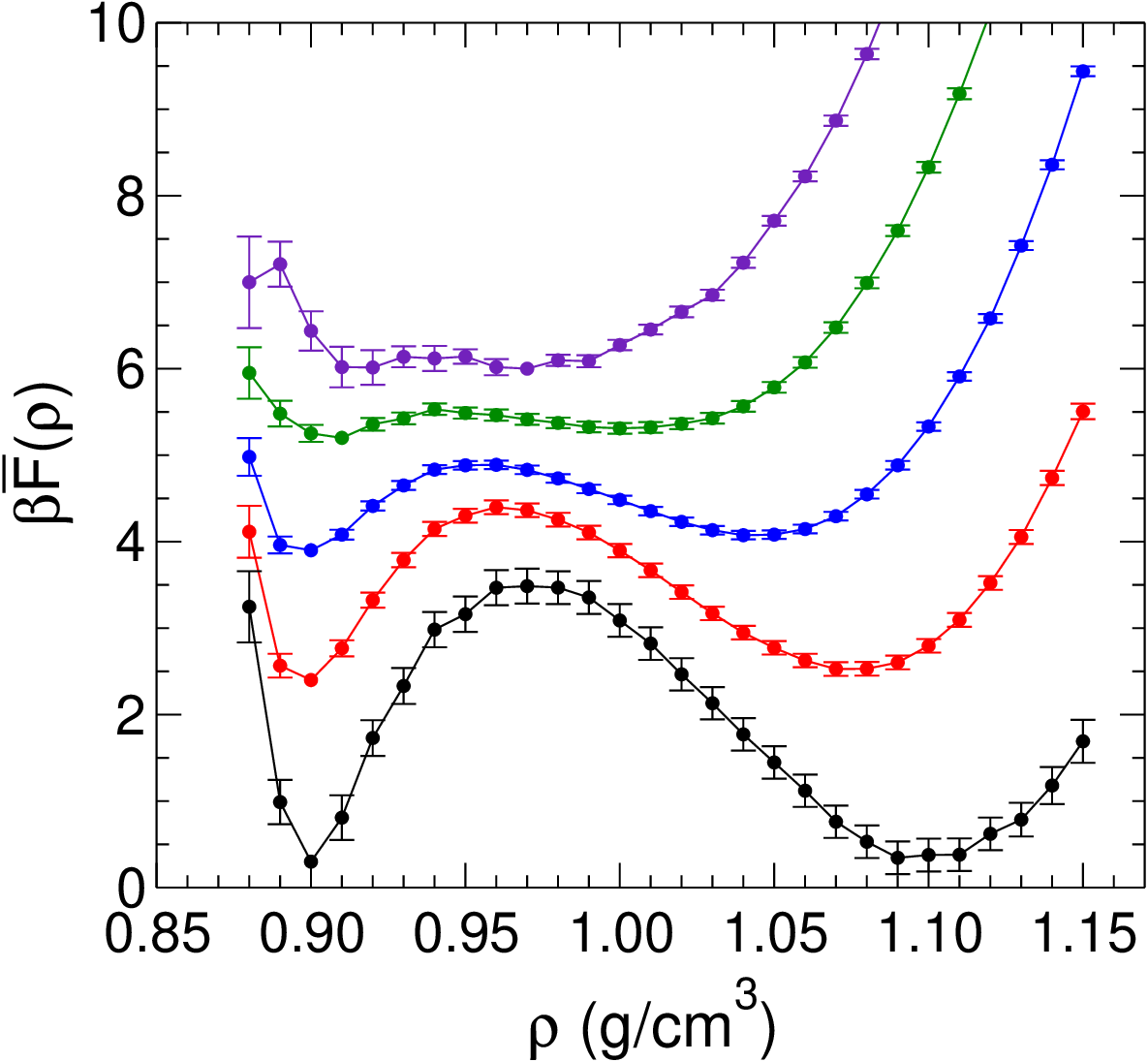}}
\caption{Contracted free energy ${\bar F}(\rho)$ at several state points approaching the liquid-liquid critical point.  From bottom to top, the state points are:
$T=230$~K and $P=245$~MPa;
$T=235$~K and $P=225$~MPa;
$T=240$~K and $P=204.5$~MPa;
$T=245$~K and $P=184$~MPa; and
$T=250$~K and $P=164$~MPa.
Each curve has been shifted by an arbitrary constant to facilitate comparison.  The combined data from all series (K, L, and M) are analyzed to obtain each curve.}
\label{vary-T}
\end{figure}

Finally, in Fig.~\ref{vary-T} we show the evolution of ${\bar F}(\rho)$ along a path in the $T$-$P$ plane that approaches the vicinity of the predicted critical point in ST2-RF.  Consistent with the occurrence of a line of first-order phase transitions terminating in a critical point, the two basins in ${\bar F}(\rho)$ are separated by a higher free energy barrier at lower $T$, which decreases in height, and then disappears, on approach to the critical point.  Fig.~\ref{vary-T} also confirms that the density of the HDL phase varies significantly with $T$, whereas that of the LDL phase is comparatively insensitive to changes in $T$.  This observation is consistent with previous studies of the free energy of ST2 water that have observed distinct HDL and LDL basins~\cite{liu1,sus,liu}.

\section{Discussion}

In summary, for the ST2-RF model, we find two distinct basins in the free energy surface $F(\rho,Q_6)$, differing in density, but both occurring at low values of $Q_6$, assuring that they correspond to disordered thermodynamic phases.  Furthermore, our results for the structural relaxation times demonstrate that both basins correspond to equilibrated metastable  liquid phases.  These observations, and the dependence of the shape and position of the basins as a function of $T$ and $P$ are entirely consistent with the occurrence of a LLPT in the ST2-RF model of water, as described in previous work~\cite{PSES,harr,denmin,sus,kessel}.  Our results are also consistent with those of Liu and coworkers for the ST2 model using an Ewald treatment of the electrostatics~\cite{liu1,liu}.  Our results are qualitatively different from the behavior of the ST2 system reported by Limmer and Chandler~\cite{lim}, and also are not consistent with their proposal that all the behavior previously ascribed to a LLPT in water-like models is in fact associated with the liquid-to-crystal transition.  

We note that Limmer and Chandler have argued that the observation of two liquid basins in $F(\rho,Q_6)$ could arise as an artifact of restricting the sampling to low values of $Q_6$;  see Fig.~9 of Ref.~\cite{supp} and the accompanying discussion.  Limmer and Chandler note that their own data for ${\bar F}(\rho)$ ``exhibits an inflection or slight minimum" for low values of $Q^{\rm max}_6$, but that this shoulder in the curve merges into the minimum associated with the crystal basin for larger values of $Q^{\rm max}_6$.  From this behavior they conclude that although the shoulder observed for small $Q^{\rm max}_6$ ``could be confused with a second liquid basin," it is in fact ``due to the barrier separating liquid from crystal."~\cite{supp}.

We disagree with this interpretation of the data.  We refer the reader to the bottom right-hand panel of Fig.~9 of Ref.~\cite{supp}, which shows the free energy surface upon which the above analysis of Limmer and Chandler is based.  In this free energy surface, the liquid basin is clearly distinct from the crystal basin, in the sense that any path connecting the minima of these two basins must pass over a barrier of at least $23kT$.  The shoulder in the free energy surface noted by Limmer and Chandler occurs deep inside the liquid basin (near $\rho=0.92$~g/cm$^3$ and $Q_6=0.08$), and is well separated from the barrier that defines the boundary between the liquid and crystal basins (near $Q_6=0.27$).  Hence, any path leading from the shoulder to the crystal basin must go ``uphill" in free energy at some point along the path.  The shoulder thus cannot be understood as an extension of the crystal basin into the low-$Q_6$ regime.  When ${\bar F}(\rho)$ is plotted for different $Q^{\rm max}_6$, the shoulder and the crystal minimum become superimposed on one another because they happen to occur at similar densities;  however, this is not a basis for concluding that these two features must be associated with the same (crystalline) free energy basin.

To conclude, we emphasize that not all models of water exhibit a LLPT.  For example, the mW model seems to be a case in which a LLPT, which otherwise might occur, becomes unobservable due to the loss of stability of the supercooled liquid with respect to crystal nucleation~\cite{sriv,mw1,lim}.  Whether or not a LLPT occurs in a given water model, and indeed in real water itself, may depend sensitively on the details of the intermolecular interaction.  For real water, it remains for experiments to determine conclusively if a LLPT can be observed, for example, by manipulating the rate of ice crystallization in the supercooled regime by exploiting nano-confinement or external fields.  
Nonetheless, the results presented here provide clear evidence that a LLPT does occur in simulations of the ST2-RF model of water, and confirm the conclusions drawn in previous studies of this model regarding the existence of a LLPT.

\section{Acknowledgements}

PHP thanks NSERC and the CRC program for support.  FS thanks ERC-PATCHYCOLLOIDS.  Computational resources were provided by ACEnet.  
We are grateful to P.G. Debenedetti, Y. Liu, J.C. Palmer, and T. Panagiotopoulos for informative discussions and for sharing their results in advance of publication.  We also thank S. McGibbon-Gardner for useful discussions.
Without implying their agreement with our methods or conclusions, we also thank D. Chandler and D. Limmer for discussions and for sharing their results in advance of publication.

\end{document}